\documentclass[iop]{emulateapj}

\def\lapp{\ifmmode\stackrel{<}{_{\sim}}\else$\stackrel{<}{_{\sim}}$\fi}
\def\gapp{\ifmmode\stackrel{>}{_{\sim}}\else$\stackrel{>}{_{\sim}}$\fi}

\newcommand{\source}{Swift~J1822.3$-$1606}
\newcommand{\src}{Swift~J1822.3$-$1606}
\newcommand{\rxte}{\textit{RXTE}}
\newcommand{\xte}{\textit{RXTE}}
\newcommand{\cxo}{\textit{Chandra}}
\newcommand{\chandra}{\textit{Chandra}}
\newcommand{\rosat}{\textit{ROSAT}}
\newcommand{\swift}{\textit{Swift}}

\newcommand{\tempo}{{\tt{TEMPO}}}

\newcommand{\degrees}{^{\circ}}
\newcommand{\kes}{PSR~J1846$-$0258}

\usepackage{amsmath}

\begin{document}

\title{The long-term post-outburst spin down and flux relaxation of magnetar Swift~J1822.3$-$1606}

\author{
P. Scholz\altaffilmark{1,2}, V. M. Kaspi\altaffilmark{1}, and A. Cumming\altaffilmark{1}
}

\altaffiltext{1}{Department of Physics, Rutherford Physics Building,
McGill University, 3600 University Street, Montreal, Quebec,
H3A 2T8, Canada}
\altaffiltext{2}{pscholz@physics.mcgill.ca}

\begin{abstract}

The magnetar \source\ entered an outburst phase in 2011 July. 
Previous X-ray studies of its post-outburst rotational evolution  
yielded inconsistent measurements of the spin-inferred magnetic field.
Here we present the timing behavior
and flux relaxation from over two years of \swift, \xte, and \cxo\ observations following the outburst.
We find that the ambiguity in previous timing solutions was due to enhanced spin down
that resembles an exponential recovery following a glitch at the outburst onset. After fitting out the effects
of the recovery, we measure a long-term spin-down rate of $\dot\nu=(-3.0\pm0.3)\times10^{-8}$\,s$^{-2}$ which implies a
dipolar magnetic field of $1.35\times10^{13}$\,G,
lower than all previous estimates for this source. We also consider the post-outburst flux evolution, 
and fit it with both empirical
and crustal cooling models. 
We discuss the flux relaxation in the context of both crustal cooling and magnetospheric relaxation
models.

\end{abstract}


\section{Introduction}

Magnetars are neutron stars whose radiation is powered by the decay of their enormous magnetic fields ($10^{13-15}$\,G)
rather than by rotation, as are the vast majority of non-accreting pulsars \citep[for reviews, see][]{wt06,mer08}. 
The large magnetic fields of magnetars
cause extreme activity such as large outbursts during which their X-ray flux output can increase
by several orders of magnitude, accompanied by short energetic X-ray bursts 
\citep[for a review of magnetar activity see][]{re11}.

Nearly every magnetar outburst has been accompanied by a change in the timing properties of the 
pulsar \citep{dk13}. The most common timing change that is observed is a spin-up glitch
contemporaneous with the outburst onset. Following a glitch, the spin period of a magnetar can
exponentially recover, although sometimes only partially, to its pre-glitch value. In some cases the recovery
has overcompensated for the spin-up glitch and the net effect is a spin down \citep{lkg10,gdk11}.
In the cases where a spin-up glitch has not been observed, other timing changes and anomalies have been
noted, such as enhanced spin down and an increase in timing noise \citep[e.g.][]{dkg09,dksg12}.

The magnetar \source, was discovered when an X-ray burst was detected by the \swift\ 
Burst Alert Telescope (BAT) on 2011 July 14 \citep{cbc+11}. It was found to be a pulsar when an $8.43$-s
periodicity was identified using \rxte\ \citep{gks11}. \citet{lsk+11} reported on initial timing
and spectroscopic results from \swift, \rxte, and \chandra\ and showed that the flux of
the source was decaying from its peak at the onset of the outburst. A spin-down rate of
$\dot{P}=2.54\times10^{-13}$ was measured 
which implied a surface dipolar magnetic field of $B=4.7\times10^{13}$\,G.
The characteristic post-outburst decay and high magnetic field value confirmed
that \source\ was a new magnetar that had experienced an outburst on 2011 July 14. 

Subsequent studies of the post-outburst timing evolution of \source\ found
that the spin-inferred magnetic field may be lower than that measured in \citet{lsk+11}. 
\citet{rie+12} presented a timing solution
in which the spin-inferred magnetic field was measured to be $2.7\times10^{13}$\,G.
\citet{snl+12} fit several timing solutions to the post-outburst \swift, \xte, and
\chandra\ observations and found that a model with only non-zero $P$ and $\dot{P}$ 
\citep[similar to what was presented in][]{rie+12}
was not a good fit to the data and that the addition of higher period derivatives improved
the fit. Their best fit utilized three significant period derivatives and the measured
$\dot{P}$ in that case implied a spin-inferred B-field of $\sim5\times10^{13}$\,G.
There has thus been some ambiguity in the best timing solution, and parameters, like $B$,
derived from those solutions for \source. 

%{\bf paragraph motivating flux evolution work}
The flux and spectral evolution of \source\ was also presented in \citet{rie+12} and 
\citet{snl+12}. They found that the flux decayed rapidly following the outburst and 
that the spectral parameters softened. This behavior followed the hardness-flux correlation 
that is generally expected in the twisted-magnetosphere model \citep{tlk02,lg06,bel09} and
observed in many other magnetars \citep[e.g.][]{icd+07,sk11}.
On the other hand, it was found that a model of the thermal relaxation of the magnetar crust
reproduced the observed luminosity decay well. Both \citet{rie+12} and \citet{snl+12}
reported that, in these models, the late time decay was not well constrained and depended on
parameters of the inner crust. 
X-ray flux observations at late times, when the source is closer to quiescence, could better constrain the
models and thus implied parameters of the neutron star crust in this picture.
The magnetospheric twist model also makes specific predictions for the flux
and spectral evolution.  In particular the twist relaxation time scale depends in
a well defined way on the stellar magnetic moment, the electric voltage sustaining
magnetospheric discharge, as well as on the emitting area, and can be tested
if these parameters are constrained \citep{bel09}.
Also, the thermal luminosity is predicted to vary as the emitting area squared, which can also be tested.

In this paper we present updated timing solutions for \source\ with a baseline that is over twice as long
as in \citet{snl+12}, using new data from the \swift\ X-ray Telescope (XRT).
From this we attempt to resolve the previous ambiguity regarding the true value of the spin-down
rate and hence the magnetic field.
We also present the up-to-date flux and spectral evolution. We discuss how the timing
and spectral evolutions have changed since previous studies and the implications 
of the new results in the
context of the magnetar model. 

%- standard magnetar intro \\
%- introduction to 1822 \\
%- summarize timing behaviour of magnetars during outbursts (glitches, timing noise etc.) \\ 
%- summarize results of \citet{lsk+11,snl+12,rie+12}. \\

\section{Observations}

\subsection{\swift\ Observations}

Since the 2011 July 14 (MJD~55756) outburst of \src, \swift/XRT \citep{bhn+05} has been used to obtain 
61 observations of the source for a total exposure time of 297\,ks. The exposure times for each individual observation
ranged from 0.5 to 18\,ks.
Data were collected in two different modes, Photon Counting (PC) and Windowed
Timing (WT). While the former gives full imaging capability with a time resolution of 2.5\,s, the latter forgoes imaging
to provide 1.76-ms time resolution by reading out events in a collapsed 
one-dimensional strip. 
As the PC mode time resolution is insufficient for our timing analysis, those data are used
only for our spectral work.

For each observation, the unfiltered Level 1 data were downloaded from the \swift\ quicklook 
archive\footnote{http://swift.gsfc.nasa.gov/cgi-bin/sdc/ql}. 
The standard XRT data reduction script, {\ttfamily xrtpipeline}, was then run using the
source position of 
RA$=18^{\rm{h}}$~$22^{\rm{m}}$\,$18^{\rm{s}}$, 
Dec$=-16\degrees$\,$04\arcmin$\,$26\farcs8$ \citep{pbk11}
and the best available spacecraft attitude file.
Events were then reduced to the solar-system barycenter using the above source position.
Source and background events were extracted using the following regions:
for WT mode, a 30-pixel long strip centered on the source was used to extract the
source events and a 50-pixel long strip positioned away from the source was
used to extract the background events. 
For PC mode, a circular region with radius 20 pixels was used for the source
region and an annulus with inner radius 40 pixels and outer radius 60 pixels was used 
as the background region. For the first (00032033001) and second (00032033017) 
PC mode observation, circular regions with radii of 6 and 2 pixels, 
respectively, were excluded to avoid pileup.

For WT mode data, exposure maps, spectra, and ancillary response files
were created for each individual orbit. The spectra and ancillary response
files were then summed to create a spectrum for each observation. For
the PC mode data, exposure maps, spectra and ancillary response files
were created on a per observation basis. We used response files for spectral fitting
from the 20130313 CALDB and version 6.13 of HEASOFT. 
In \swift/XRT observations, there are columns of bad pixels that can disrupt
the source PSF.
Orbits were not used in an observation if the bad columns were found to be 
within 3 pixels of the source position. 

\subsection{\rxte\ and \cxo\ Observations} 
\label{sec:rxteobs}
We downloaded 32 \rxte\ observations from the {\em HEASARC} archive. 
These data spanned the MJD range from 55758 to 55893 (2011 July 16 to 2011 Nov 28),
for a total of 174\,ks of integration time.
The data were collected in {\tt GoodXenon} mode which records each event with
1-$\mu$s time resolution. 

%We selected events in the 2--10 keV energy range (PCA channels 6--14) from the top xenon layer of each PCU 
%for our analysis, to maximize signal-to-noise ratio. 
%The data from all the active PCUs were then merged. If more than one
%observation occurred in a 24-hr period, the observations were combined into a single data set.
%Photon arrival times were adjusted to the solar system barycenter using the same position
%as the for \swift\ data. Events were then binned into time series with resolution 1/32 s for use
%in the following analysis.

Following the outburst, our ToO program with the \emph{Chandra
X-ray Observatory} was triggered.
Five ACIS Continuous Clocking (CC) mode observations were obtained between MJD 55769 and 56036 
(2011 July 27 and 2012 April 19), with exposures ranging from 10 to 20\,ks. 
CC-mode has a time resolution of 2.85\,ms
and sensitivity between 0.3 and 10\,keV\footnote{\url{http://cxc.harvard.edu/proposer/POG/html/}}.

The \rxte\ and \cxo\ sets of observations are identical to those used in 
\citet{snl+12} and are summarized in Table 1 of that work. 

\section{Analysis \& Results}

\subsection{Timing}

For each \swift\ and \chandra\ observation, a pulse time of arrival was extracted using the maximum likelihood method
described in \citet{snl+12}. \xte\ TOAs were measured using cross-correlation with a template profile, as
the maximum likelihood method is computationally too expensive for these data due to the high number of counts. 
We then fit timing solutions to the TOAs using the \tempo\footnote{\url{http://tempo.sourceforge.net}} pulsar timing
software package. 

\begin{figure}
\plotone{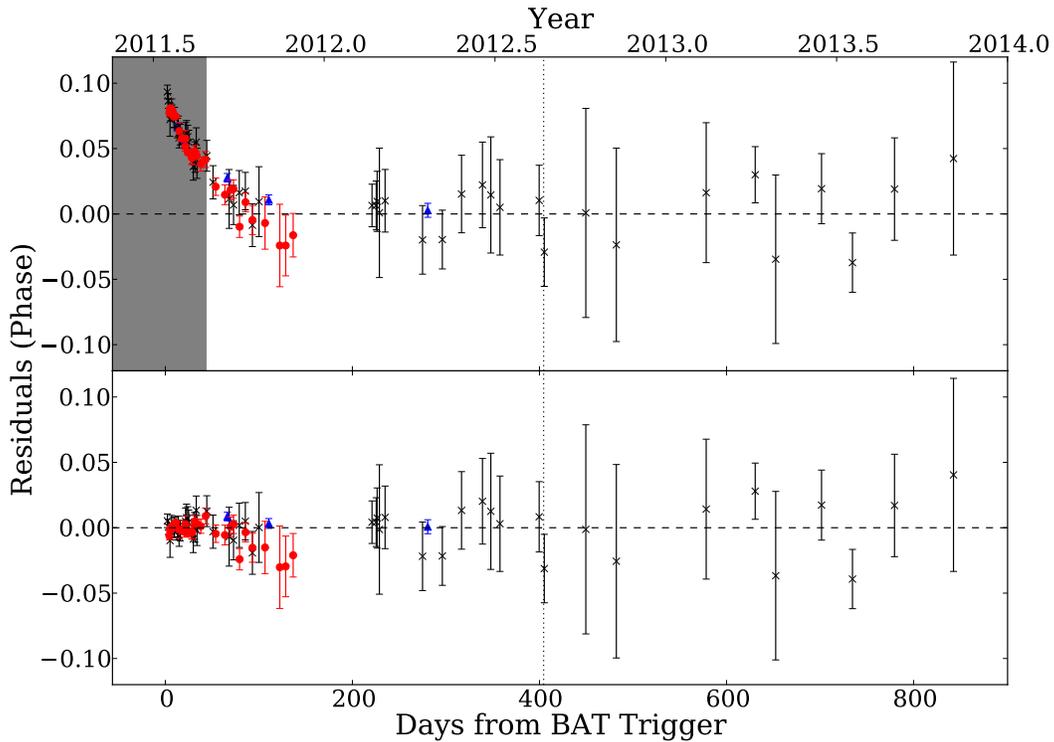}
\figcaption{
Timing residuals of \src. The top panel shows the solution in Table \ref{ta:timing} before the glitch recovery is fit.
The bottom panel shows the residuals for the solution with the glitch recovery fit out.
In both panels, black crosses represent \swift\ observations, red circles indicate \rxte\ observations, and \cxo\ data
are shown as blue triangles.
The gray band represents the data that were excluded when the long-term spin down was fit prior to attempting to fit
the glitch recovery.
The vertical dotted line shows where the longest previous published timing solution ended \citep{snl+12}.
\label{fig:resids}
}
\end{figure}

We first fit a timing solution that included only a frequency and frequency derivative as was done by \citet{rie+12} 
and for Solution 1 of \citet{snl+12}. This model did not fit the data well, with reduced $\chi^2_\nu=7.38$ for 
83 degrees of freedom.
In order to be sensitive to only
the long-term spin down of the pulsar, 
we then fit the same model to all TOAs from observations 
taken later than approximately two months (MJD~$\ge55800$) 
from the onset of the outburst.
We found that this provided an acceptable fit ($\chi^2_\nu/\nu=1.19/41$), and that
it was much improved from the $\nu$ and $\dot\nu$ fit to the entire data set.
We noted
that the excluded TOAs appeared to form an exponential decay in the phase residuals (top-panel of Figure \ref{fig:resids}). 
We therefore added an exponential
glitch recovery to our model with the glitch epoch at the time of the \swift/BAT burst trigger (MJD~55756).
The glitch recovery model provided an excellent fit with $\chi^2_\nu/\nu=0.97/81$.
Table \ref{ta:timing} shows the results of this fit and the bottom panel of Figure \ref{fig:resids} shows the best-fit residuals.

In addition to the exponential recovery model, we tried some alternative models. We attempted to fit the residuals with higher
derivatives similar to \citet{snl+12}. We found that in order to produce a fit with a similar $\chi^2_\nu$, 
five frequency derivatives were needed. 
However, since the long-term spin down is well fit by the simpler $\nu$ and $\dot\nu$ timing solution,
and extra derivatives are not needed if the first two months following the outburst is ignored,
it is clear that the multi-derivative solution is not representative of the true spin down of the 
pulsar and was only an artifact due to the contamination of the enhanced spin down at early times.
We also compared the two models by fitting them to all of the data excluding the last three months
and seeing how well they predicted the excluded TOAs. The single-derivative glitch recovery model
predicted the last three months within 5\% in phase and the measured parameters were fully consistent with
the fit to the whole data set. The multi-derivative model, however, did not predict the later data with
the last three months of TOAs wandering up to 30\% in phase.

We also attempted to fit the post-outburst TOAs
with a change in $\nu$ at MJD~55900 as the timing residuals indicated a deviation in the rotation from the long-term
spin down prior to that epoch. 
This did not provide an acceptable fit, so we also tried a change in $\dot\nu$. This 
gave a slightly higher $\chi^2_\nu/\nu$ (1.22/81) than in the exponential recovery model. 
Aside from the poorer fit, this latter solution seems contrived given the absence of precedence for such behavior
post-outburst in magnetars. As discussed in Section \ref{sec:timing_disc}, there are many examples of glitches
accompanied by exponential recoveries in magnetar outbursts. For these and the aforementioned reasons, 
we conclude that the glitch recovery model is by far the most likely.

Although we model the post-glitch spin down with a glitch recovery, we cannot conclusively say that a glitch occurred. 
This is because timing observations of the source pre-outburst are not available.
However, if the pre-detection spin-down frequency was the same as its long-term post-outburst value 
(i.e. the exponential recovery perfectly compensated for the spin-up glitch),
the fractional magnitude of the hypothetical glitch would be $\Delta\nu/\nu = (2.3\pm0.1)\times10^{-7}$ which is in the typical
range of glitch magnitudes observed from magnetars \citep[$\Delta\nu/\nu\sim10^{-7}-10^{-5}$; e.g.][]{dk13}.
An under-recovery wound imply a larger glitch, whereas an over-recovery a smaller glitch.

%- extracted TOAs using ML method \\
%- to get long-term spin-down, fit $\nu$ and $\dot{\nu}$ model to TOAs using TEMPO while excluding first two months of data \\
%- found above model fit well \\
%- reincluded first two months and saw that it looked like exponential decay \\
%- fit with glitch at burst epoch with a $\Delta\nu\exp({-t/\tau_d})$ \\
%- since we don't have any data from before the onset of the outburst, can't say whether it was a glitch or just enhanced spin-down.
%- if we assume that the pre-outburst spin-down was exactly like the post-recovery spin-down (unlikely to be exact), then the glitch
%was $\Delta\nu/\nu=X.X\times10^{-7}$, which is a typical magnetar glitch size. \\

\subsection{Flux and Spectral Evolution}
\label{sec:flux_analysis}

We fit a photoelectrically absorbed blackbody plus power-law model to each \swift\ and \chandra\ spectrum
using XSPEC\footnote{\url{http://xspec.gfsc.nasa.gov}} v12.8. The hydrogen column density, $N_\mathrm{H}$,
was fixed to the value measured by \citet{snl+12}, $4.53\times10^{21}$\,cm$^{-2}$, and we used the XSPEC {\tt phabs} model
with abundances from \citet{ag89} and photoelectric cross sections from \citet{bm92}.
For observations later than MJD~55975, sets of observations nearby in time 
were fit with joint $kT$ and $\Gamma$. This was done as the spectral parameters were not well constrained for individual
observations and for each set $kT$ and $\Gamma$ were consistent from observation to observation.
The flux was left free to vary from observation to observation as significant flux evolution was still present.
The reduced $\chi^2_\nu$ values from the spectral fits ranged from 0.81 to 1.5. 
%{\bf present $\chi^2$ values with specral params in an table? (probably too big)}.

Figure \ref{fig:spec} shows the results of the spectral fits. The flux is seen to decay following the outburst
and the $kT$ and $\Gamma$ spectral parameters soften. The blackbody temperature, $kT$, remained approximately 
constant at $\sim 0.75$\,keV, 
or perhaps even increased, in the first 10 days following the outburst onset at the BAT trigger.
It then decreased to $\sim0.6$\,keV between 10--100 days from the trigger. Since MJD~55900, the 
blackbody temperature has remained roughly constant. The photon-index, $\Gamma$ appears to have
increased (softened) following the outburst.
This is most evident from the \chandra\ observations where the photon-index softened from $\sim2.0$ to $\sim2.5$ between
10 and 300 days from the trigger.

In order to characterize the flux relaxation we fit exponential decay models to the fluxes measured from the X-ray spectra.
We first attempted a double-exponential decay model, $F(t)=F_1\exp^{-(t-t_0)/\tau_1}+F_2\exp^{-(t-t_0)/\tau_2}+F_q$. 
This did not provide an acceptable
fit, as the reduced $\chi^2_\nu/\nu$ was 2.26/56. So, we then fitted a triple exponential model: 
$F(t)=F_1\exp^{-(t-t_0)/\tau_1}+F_2\exp^{-(t-t_0)/\tau_2}+F_3\exp^{-(t-t_0)/\tau_3}+F_q$.
In both cases, $F_q$ is the quiescent 1--10\,keV flux, $3\times10^{-14}$\,erg\,cm$^{-2}$\,s$^{-1}$,
implied from the 0.2--2.4 keV flux and spectral model measured from a 1993 \rosat\ observation \citep{snl+12}.
The triple exponential model provides a much better, and acceptable, fit with $\chi^2_\nu/\nu=0.95/54$.
The best-fit exponential decay timescales were $6\pm1$, $27\pm2$, and $320\pm20$ days.
%This third, slower, decay timescale shows that the flux of \source\ is decreasing at a slower rate than predicted
%by the double-exponential fit in \citet{snl+12}.

%- spectra fit in Xspec \\
%- for later spectra when count-rate is lower, $kT$ and $\Gamma$ were tied for nearby observations in time. The flux stayed free for all observations \\
%- present best-fit model to flux decay (this time with 3 exponentials!). i.e. slower than predicted by \citet{snl+12} \\
%- kT decreasing and Gamma increasing as in \citet{snl+12} \\
%- andrew's model fit to data? \\

\section{Discussion}

We have presented over a year of new \swift/XRT observations of \source\ which have increased the baseline to over two and a half years.
This has allowed us to better measure
the timing and spectral evolution. Importantly, we have presented a more straightforward timing model than that in \citet{snl+12}
as well as evidence for a glitch at the epoch of the outburst onset.
Our new timing model implies a long-term spin-down rate that is significantly smaller than previous estimates.
We have also updated the post-outburst flux and spectral evolution of \source\ and show that it is well fit by a triple-exponential
model that is decaying to the \rosat\ measured quiescent flux.
Below we discuss the implications of these findings.

\subsection{Post-outburst spin-down behavior}
\label{sec:timing_disc}

Previous studies of the timing evolution of \source\ did not find a consistent timing solution. \citet{lsk+11} first presented
a timing solution with just $\nu$ and $\dot\nu$ fit to the first $\sim80$ days following the outburst and measured 
$B=4.7\times10^{13}$\,G. This was followed by \citet{rie+12} who, with 275 days of observations, found
a lower value of $B=2.7\times10^{13}$\,G. Both studies found that there was an unmodelled trend
in their residuals, possibly attributed to timing noise. \citet{snl+12} attempted to fit the trend with higher frequency
derivatives. They found that as higher derivatives were added, the value of $B$ increased. The best fit
was found with three frequency derivatives (Solution 3) and $B=5.1\times10^{13}$\,G, larger than the \citet{rie+12} estimate. 

Here we find that in the initial $\sim100$ days following the onset of the outburst, the timing behavior of \source\ was not
representative of the long-term spin down of the pulsar. Specifically, it was spinning down more rapidly, likely due to a recovery
from a glitch. By using the data from the initial post-outburst epochs, the previous studies measured higher spin-down rates because of contamination
from the early enhanced spin down. This is shown by the fact, pointed out by \citet{tx13}, that as the solutions were derived from
longer timing baselines, the measured spin-down rate became lower.

Glitches in pulsars and magnetars have been known to show exponential recoveries. 
Magnetars 1RXS~J170849.0$-$400910 and 1E~2259+586 showed clear exponential 
recoveries following their 2001 and 2002 glitches \citep{kg03,wkt+04}
and 4U~0142+61 showed a slight over-recovery following its 2006 glitch \citep{gdk11}. 
The magnetically active
rotation-powered pulsar \kes\ showed a spin-up glitch with a large over-recovery \citep{lkg10}.
The decay timescale measured for \source\ (40 days; see Table \ref{ta:timing}) 
is comparable to the analogous timescale for the 2001 glitch of 1RXS~J170849.0$-$400910 (43 days)
and somewhat longer than was observed in 1E~2259+586 (17 days) and 4U~0142+61 \citep[12--17 days;][]{dk13}.
We note that in the single well sampled anti-glitch, during the 2012 outburst of magnetar 1E~2259+586, 
no exponential recovery was seen \citep{akn+13}.
Without knowledge of the pre-outburst
spin down of \source, we cannot conclusively argue for the occurrence of a glitch, but given the exponential form 
of the post-outburst spin down
and the prominence of exponential glitch recoveries in other magnetars, it seems extremely plausible. 

However, the occurrence of enhanced spin down without the occurrence of a glitch following a magnetar outburst has also been observed. 
Following the the 2002 outburst of 1E~1048.1$-$5937 no glitch was observed and the magnitude of $\dot\nu$ increased as the pulsed flux decreased \citep{dkg09}.
The magnetars 1E~1547.0$-$5408 and SGR~1745$-$2900 showed similar behavior after their 2008 and 2013 outbursts, respectively \citep{dksg12,kab+13}. 
The same behavior is clearly not present in the post-outburst
timing of \source. In its case, as the flux of the magnetar decreased, the magnitude of $\dot\nu$ decayed. 
Thus, the post-outburst timing behavior of \source\ looks like a glitch recovery as observed from 1RXS~J170849.0$-$400910, 1E~2259+586,
and 4U~0142+61 and does not resemble the more unusual enhanced spin-down behavior
seen in 1E~1048.1$-$5937 and 1E~1547.0$-$5408.

The enhanced spin-down rate following the outburst of \source\ was a factor of several larger than its long-term spin-down rate. 
The instantaneous post-glitch spin-down rate at the glitch epoch for an exponential recovery 
can be quantified by $\Delta\nu_d/\tau$ \citep{dkg08}. For \source, this
quantity is $\Delta\nu_d/\tau=(26\pm5)\dot\nu$ where $\dot{\nu}$ is the value we report in Table 1. 
This is higher than in previous magnetar glitches, the next highest being
the 2002 glitch of 1E~2259+586 with a $\Delta\nu_d/\tau=(8.2\pm0.6)\dot\nu$.
For radio pulsars, the post-glitch recoveries usually result in spin-down enhancements of only a few percent 
\citep[e.g.][]{fla90,wbl01}.

Since many magnetars are observed frequently after their outburst, but are not followed up with long-term timing campaigns
as has been \source, it is possible that the measurements of their spin-down rates are not representative of their long-term timing
evolution. 
%If the magnetic field of a magnetar is measured based on incoherent period measurements shortly
%after its outburst, or based on a short-term coherent timing solution, the value measured may be different from
%the true long-term spin-down magnetic field. 
For example, the magnetic field of CXOU~J164710.2$-$455216 was measured
to be $\sim1\times10^{14}$\,G using phase-coherent timing in the first 100-200 days following its 2006 outburst \citep{icd+07,wkga11},
but \citet{akac13} have recently placed an upper limit on the long-term spin-down rate and hence inferred magnetic field 
of  $<7\times10^{13}$\,G using a timing baseline of $\sim6$ years.

The magnetic field measured from the long-term spin down of \source\ is $1.35\times10^{13}$\,G. This is the second
lowest spin-down inferred magnetar dipolar magnetic field and is about a factor of two higher than the lowest measured value,
$6.1\times10^{12}$\,G of SGR~0418+5729 \citep{rip+13}. 
However, since pulsar timing is only sensitive to the surface dipolar component of the field, 
the true magnetic field of the magnetar could be significantly higher than the value
measured from pulsar timing if the toroidal component is much higher than the poloidal component
or if the polodial field has significant multipolar contributions.
Indeed, there is evidence for a much higher magnetic field for SGR~0418+5729 from spectral modelling \citep[$\sim1\times10^{14}$\,G,][]{ggo11}
and from a phase-resolved cyclotron absorption feature \citep[$>2\times10^{14}$\,G,][]{tem+13}.

%- discuss what \citet{rie+12} and \citet{snl+12} found, and why they didn't get the right answer. \\
%- mention that Tong \& Xu (2013) since they `reported' a decreasing period derivative \\
%- discuss occurrence of 'enhanced spin-down' after magnetar outbursts (1547, others) \\
%- say that measurements of timing parameters after an outburst may not be representative of the long-term spin-down $\rightarrow$ B-field of some magnetars may be overestimated \\

\subsection{Models of flux relaxation}

In addition to the empirical model fits in Section \ref{sec:flux_analysis}, 
we fit the light curve with a model of crustal cooling (Cumming et al., in prep). 
In these models, the neutron star crust is heated by a sudden deposition of energy, and the subsequent thermal relaxation 
and cooling is followed by integrating the thermal diffusion equation in time. Both \citet{snl+12} and \citet{rie+12} 
were able to fit the first 100 days of the light curve by depositing $\sim 10^{42}\ {\rm erg}$ in the outer crust 
at densities $\sim 10^9$--$10^{10}\ {\rm g\ cm^{-3}}$. \citet{snl+12} found that at later times the model 
did not fit the data well, declining in flux more rapidly than observed. 
However, the late time lightcurve is sensitive to a number of physics inputs such as the neutron contribution 
to the heat capacity near neutron drip, thermal conductivity of the inner crust, and angular distribution of the 
heating around the star. We investigate this further here, updating our cooling models, 
and including the new flux measurements at times $\gtrsim 500$ days after outburst. 

We have made two improvements to the model shown in \citet{snl+12}. The first is in the treatment of the outer envelope. 
Rather than use the analytic flux-temperature relations for the outer envelope from \citet{py01}, 
which are strictly correct only in the high density isothermal part of the crust, 
we have reproduced the detailed magnetized envelope models of \citet{py01}, 
allowing us to match the envelope to the correct density at the top of our numerical grid 
(we typically follow the temperature on a grid in density extending from the crust/core boundary 
to $\approx 6\times 10^8\ {\rm g\ cm^{-2}}$).  
Because the temperature profile is not fully isothermal at these depths 
(deeper in the crust, the temperature profile is close to isothermal in steady-state because of the large thermal conductivity),  
using the analytic flux-temperature relation leads to a tens of percent underestimate of the flux and corresponding 
overestimate of the energy deposited. Second, we have improved our treatment of the magnetic field geometry. 
We calculate the cooling curve at the magnetic pole, where the magnetic field is radial, 
and then rescale that light curve appropriately for each local patch on the neutron star surface given the 
local magnetic field direction (a time-dependent extension of \citet{gh83}). 
For heating across the whole surface and a dipole magnetic field, 
we find a more extended tail of the light curve at times of several hundred days and onwards compared to \citet{snl+12}; 
the shape of the early time light curve does not change significantly. 
The late time tail comes from the equatorial region in which the field is close to horizontal, 
significantly reducing the radial thermal conductivity and increasing the cooling time.

\begin{figure}
\plotone{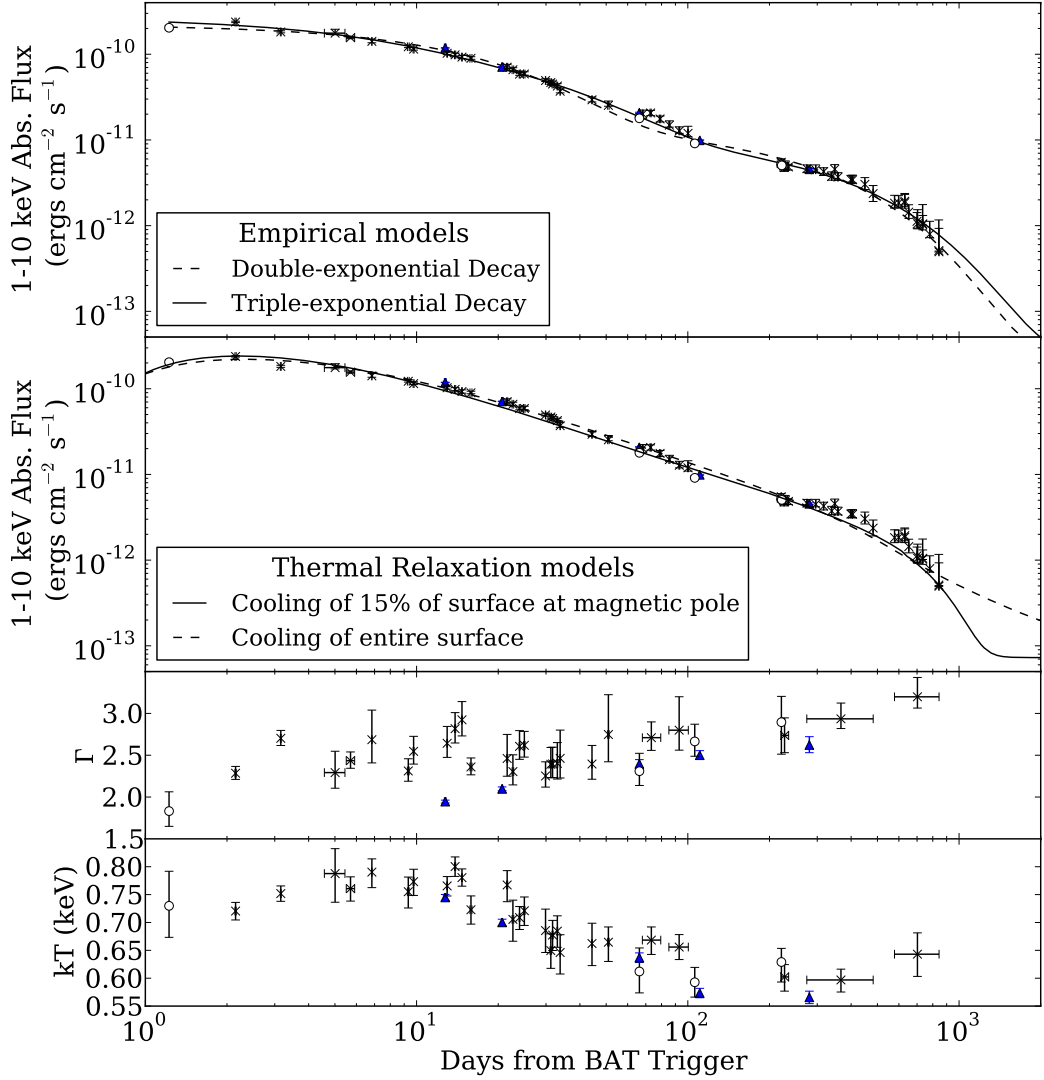}
\figcaption{
Spectral evolution of \src. The top two panels both show the 1--10\,keV flux evolution, but are fitted with two different
sets of models.
Black crosses denote \swift\ WT mode observations, open circles represent \swift\ PC mode data,
and blue triangles show \chandra\ observations. In the spectral fits, for some observations $kT$ and $\Gamma$ are fit jointly in sets
that are nearby in time, and so are represented by a single point in their respective plots.
The horizontal error bars on $kT$ and $\Gamma$ respresent the extent in time of such sets.
\label{fig:spec}
}
\end{figure}

Two models that reproduce the first 100 days of the light curve are shown in the second panel of Figure \ref{fig:spec}. 
We assume a distance of 1.6 kpc, neutron star mass $1.6\ M_\odot$, radius $R=11.2\ {\rm km}$, 
core temperature $1.5\times 10^7\ {\rm K}$, and magnetic field strength at the pole $10^{14}\ {\rm G}$. 
The mass and radius correspond to the particular equation of state chosen by \citet{bc09}. 
As discussed in \citet{snl+12}, a different choice of mass and radius primarily changes the cooling timescale 
by a factor of $1/g^2$, where $g$ is the gravity, because the crust thickness is $\propto 1/g$. 
The core temperature is chosen so that the luminosity at late times is $2\times 10^{31}\ {\rm erg\ s^{-1}}$, 
comparable to the \rosat\ measurement of the quiescent flux.  
We then vary the energy injected and depth of heating to match the lightcurve. 
In the models shown in Figure \ref{fig:spec}, we deposit $\approx 3\times 10^{42}\ {\rm erg}$ 
in the outer crust down to a density of $\approx 10^{11}\ {\rm g\ cm^{-3}}$, 
either over the whole surface assuming a dipole field geometry \citep[as in][]{snl+12}, 
or over 15\% of the neutron star surface at the magnetic pole. 
Both models predict the same flux evolution for times $\lesssim 100$ days, but differ at late times. 
Heating a small region near the pole gives better agreement with the shape of the late time light curve. 
In this case the field is close to radial everywhere in the heated region, 
leading to more rapid cooling at late times which more closely matches the observed luminosity decrease. 

Both models underpredict the luminosity at times $\gtrsim 200$ days by about a factor of two. As emphasized by \citet{snl+12} 
and \citet{rie+12} there are several properties of the inner crust that can change the late time part of the light curve. 
A small amount of heat deposited in the inner crust can bring the late time lightcurve into agreement with the data. 
The shape of the decay at times $\gtrsim 200$ days is sensitive to the heat capacity of the layer of
normal neutrons just below neutron drip. 
The thickness of this layer is determined by how quickly the critical temperature $T_c$ for superfluidity 
increases with density \citep{pr12a}. We find that the best fits are obtained when we do not include 
the neutron heat capacity at all, i.e.~$T_c$ rises very rapidly with density below neutron drip; 
a slow rise in $T_c$ leads to a slower decline in the light curve at late times. 
Another important parameter is the impurity parameter $Q_{\rm imp}$ in the inner crust 
which determines the thermal conductivity. We set $Q_{\rm imp}\approx 10$ in the models shown in Figure \ref{fig:spec}.

It is rather remarkable that the models we have computed with standard assumptions for the inner crust physics reproduce the 
shape of the observed decay so well. It would be interesting to carry out a comprehensive survey of the parameters such as gravity, 
critical temperature for neutron superfluidity, inner crust thermal conductivity, magnetic field geometry and 
angular size of the heated region \citep[e.g.~see the similar study by][for the accreting neutron star XTE~J1701$-$462]{pr13}. 
The good agreement is especially interesting given the uncertainties in comparing the models with the data. 
These include the fact that the models predict the surface bolometric luminosity of the star, 
whereas the X-ray flux is measured in the narrow bandpass $1$--$10\ {\rm keV}$. 
Similarly, the spectrum of the emission is not predominantly a thermal spectrum but instead is dominated by a power law component. 
Furthermore, the thermal part of the spectrum that is observed decreases in flux primarily due to a decreasing 
emitting area rather than a decreasing temperature as might be expected for a cooling surface. 
Explaining the observed decrease in emitting area is the major challenge for crust cooling models.
Addressing the spectral evolution requires detailed modelling of the formation of the spectrum. 
Here we have assumed that while there can be significant modification of the spectrum, 
due to for example resonant Compton scattering in the magnetosphere, the luminosity remains largely unaffected,
at least over the bulk of the evolution. 

The magnetospheric untwisting model detailed by \citet{bel09} also makes specific
predictions regarding the flux and spectral evolution of magnetars post-outburst.
In this picture the post-outburst magnetosphere has been twisted due to crustal motions originating
from stresses induced by the strong internal magnetic field.  The twist is carried
by a bundle of current-carrying field lines (the ``j-bundle'') which is anchored in
the crust on a footpoint of area $A$.  This footpoint, bombarded by current particles,
radiates thermal emission and both fades and shrinks as the j-bundle dissipates
during relaxation and untwisting.  The evolution time of the X-ray luminosity is
predicted to be 
$t_{ev} \simeq 10^7 \mu_{32} \Phi_{10}^{-1} A_{11.5}$~s \citep{bel09,mgz+13},
where $\mu_{32}$ is the magnetic moment in units of $10^{32}$~G~cm$^{3}$,
$\Phi_{10}^{-1}$ is the electric voltage sustaining $e^{\pm}$ discharge in the magnetosphere
in units of $10^{10}$~V,
and $A_{11.5}$ is the j-bundle footpoint area in units of $10^{11.5}$~cm$^2$.
For \source, the relation predicts an evolution
time scale of $\sim 10^6$ s, or $\sim10$ days assuming
$\Phi_{10}^{-1} = 1$, reasonable given expectations \citep{bel09}.  
This is roughly consistent with the time scale ($\sim6$ days)
we found for the fastest-decaying exponential component (see Section~\ref{sec:flux_analysis})
but inconsistent with the second two time scales in the three-component
decay model, unless $\Phi_{10}$ is significantly smaller than unity.  
It could be that the j-bundle untwisting time scale corresponds to
the shortest exponential decay, with overall crustal cooling corresponding to the latter two.

\begin{figure}
\plotone{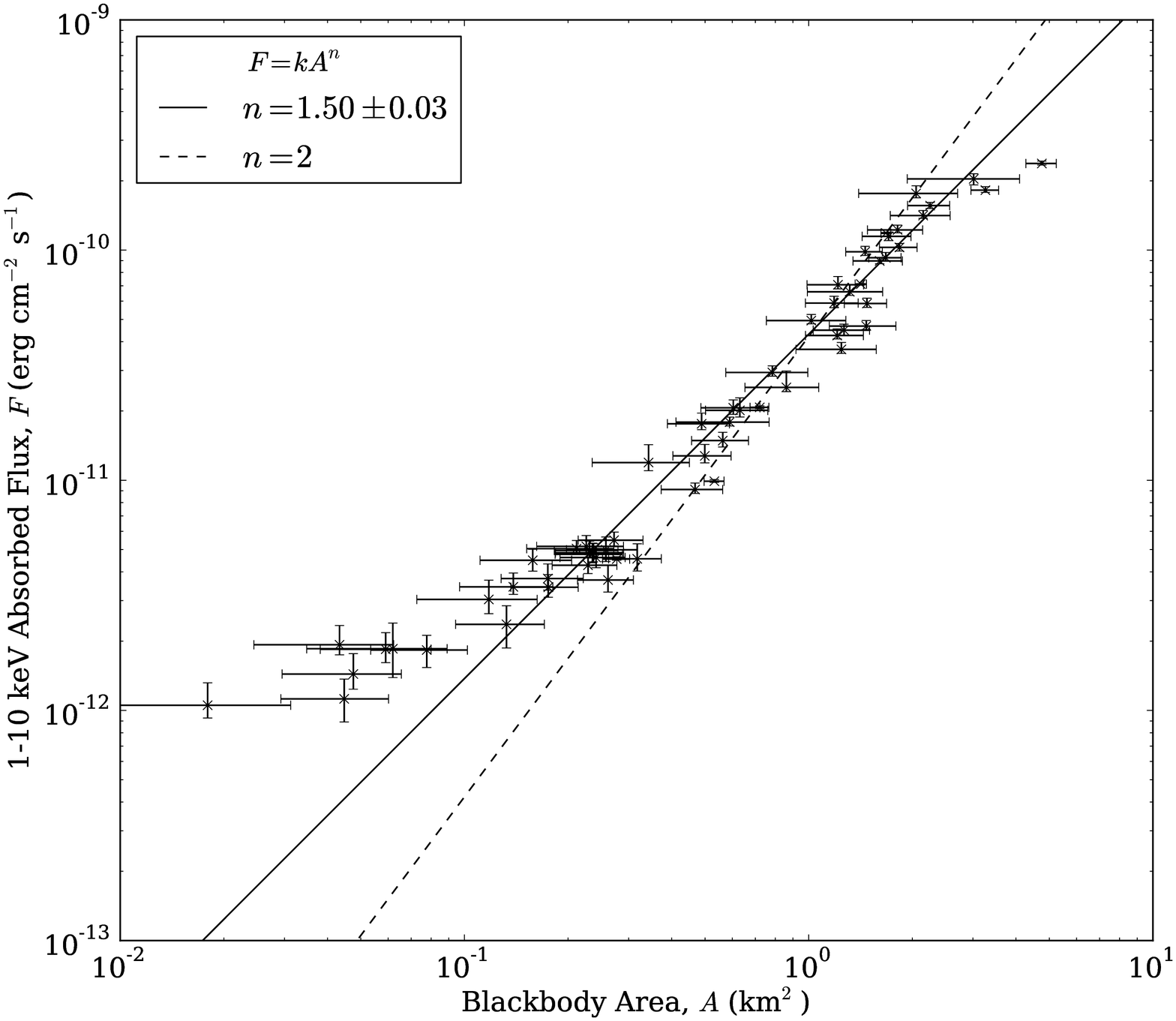}
\figcaption{
1--10\,keV absorbed flux as a function of the emitting area of the blackbody component of the spectral fits.
To these data, we have fit the function $F=kA^n$. The solid line shows our best fit when allowing $n$ to vary, 
The best fit line in that case, in terms of luminosity (assuming a distance of 1.6\,kpc), is
$L=1.5\times10^{34}{\rm\,erg\,s^{-1}}(A/1\,{\rm km^2})^{3/2}$. The dotted
line shows the case $n=2$, as predicted by \citet{bel09}. For that case, the best-fit line is given
by $L=1.3\times10^{34}{\rm\,erg\,s^{-1}}(A/1\,{\rm km^2})^2$.
\label{fig:bbarea}
}
\end{figure}

The untwisting model further predicts that the thermal X-ray luminosity from the heated footpoint 
should vary with $A^2$. Figure \ref{fig:bbarea} shows the thermal X-ray flux of \source\ 
plotted against the emitting area as inferred from our blackbody fits. 
Here we compare the expectation of the magnetospheric twist model, 
namely $f\propto A^2$ with our data and find that this is not a good description of the data. 
Rather, our best-fit relation has $f\propto A^{3/2}$. 
The observed prefactor is larger than the model predicts even when the twist is maximal (of order unity) 
and the voltage drop is $\sim10\ {\rm GeV}$ ($\Phi_{10}\approx 1$). 
\citet{bel09} predicts $L\approx 10^{34}\ {\rm erg\ s^{-1}}\ B_{14}R_6^{-3}\psi \Phi_{10} (A/{\rm km^2})^2$, 
where $\psi$ is the twist angle. 
The observed relation has $L=1.5\times 10^{34}\ {\rm erg\ s^{-1}}$ for $A=1\ {\rm km^2}$ and $d=1.6\ {\rm kpc}$. 
However, $\Phi_{10}=1$ is on the upper end of the $0.1-1\,\Phi_{10}$ range set by pair creation given by 
\citet{bel09}, and the magnetic field we infer from spin down is significantly smaller than $10^{14}$\,G, 
reducing the expected luminosity. In addition, the absorbed 1--10 keV flux is an underestimate of the 
true bolometric flux.

The shallower scaling of luminosity with area could arise from a systematic change in the shape of the
thermal spectrum compared to a blackbody as the flux decreases. 
Then the inferred area from blackbody fits would systematically change with flux, 
modifying the underlying $L\propto A^2$ scaling. 
An alternative physical explanation is that the field geometry is more complex than the 
dipole geometry assumed by \citet{bel09}.
The luminosity is $L\approx I{\mathcal V}$ where $I$ is the current, 
given from Amp\`ere's law by $I\propto B_\phi a$ where $a$ is the radius of the flux bundle 
at the surface of the star. Following the bundle of field lines from one pole to the other 
through the magnetosphere \citep[see][Appendix A for a more detailed treatment]{bel09}, 
the twist angle is $\psi\approx (r_{\rm max}/a)(B_\phi/B_P)$, 
where $B_P$ is the poloidal field strength, $B_\phi$ the toroidal field in the twist, 
and $r_{\rm max}$ is the maximum radial extent of the flux tube. 
For a dipole field, $r_{\rm max}\approx R(R/a)^2$, giving $L\propto a^4\propto A^2$
as found by \citet{bel09}. 
A different field geometry changes the scaling. 
For example, a quadrupole field has $r_{\rm max}\approx R(R/a)$, giving $L\propto a^3\propto A^{3/2}$, 
in agreement with the observed scaling in Figure \ref{fig:bbarea}.
It may therefore be of interest to explore magnetospheric untwisting with a more complex field geometry.

\section{Conclusions}

We have presented an up-to-date analysis of the post-outburst flux and timing evolution for \source. 
We find that the spin down following the outburst is well described with an exponential glitch recovery
and that the long-term spin-down inferred magnetic field is lower than previously estimated. 
From this, we conclude that a glitch likely occurred near the outburst onset as has been seen in several
other magnetar outbursts. We also find that the post-outburst flux evolution is consistent with 
thermal relaxation of the neutron star crust, particularly if heat was deposited internally in 
a small region close to neutron drip depth and near the magnetic axis. We find that flux relaxation due to magnetospheric 
untwisting may also be consistent if the poloidal magnetic field is more complicated than
a simple dipole. \\

We are grateful to the \swift, \chandra, and \rxte\ teams for their flexibility
in scheduling TOO observations. We thank Kostas Gourgouliatos for useful discussions. 
VMK holds the Lorne Trottier Chair in Astrophysics and Cosmology
and a Canadian Research Chair in Observational Astrophysics and received additional support from 
NSERC via a Discovery Grant and
Accelerator Supplement, by FQRNT via the Centre de Recherche Astrophysique de Quebec, 
and by the Canadian Institute for Advanced Research.

\bibliographystyle{apj}
\bibliography{journals_apj,/homes/borgii/pscholz/Documents/papers/myrefs,modrefs,psrrefs,crossrefs,/homes/janeway/maggie/Tex/maggie_refs}
%\bibliography{journals_apj,myrefs,modrefs,psrrefs,crossrefs,/homes/janeway/maggie/Tex/maggie_refs}
%\bibliography{journals1,myrefs,modrefs,psrrefs,crossrefs}

\begin{deluxetable}{lc}
\tablecaption{Timing Parameters for \src.
\label{ta:timing}}
\tablewidth{0pt}
\tablehead{
\colhead{Parameter}&\colhead{Value}
}
\startdata
Observation Dates & 16 July 2011 - 04 Nov 2013 \\
Dates (MJD) & $55758 - 56600$\\
Epoch (MJD) & $55 761$\\
Number of TOAs & $86$\\
$\nu$  (s$^{-1}$)  & $0.118 515 4135(9)$\\
$\dot{\nu}$ (s$^{-2}$) & $-3.0(3) \times 10^{-16}$  \\
\hline
\multicolumn{2}{c}{Post-Outburst Glitch Recovery} \\
\hline
Glitch (Burst) Epoch (MJD)  & $56756.000$\\ 
$\Delta \nu_d$ (s$^{-1}$)   & $2.7(1)  \times 10^{-8}$\\
$\tau_d$ (days) & $40(6)$ \\ 
RMS residuals (ms)  & $24.3$\\
$\chi^2_\nu / \nu$  & $0.97/81$\\
\hline
\multicolumn{2}{c}{Derived Parameters} \\
\hline
$B$ (G) & $1.35(6)\times10^{13}$ \\
$\dot{E}$ (erg\,s$^{-1}$) & $1.4(1)\times10^{30}$ \\
$\tau_c$ (kyr) & 6300(600)
\enddata \\
Numbers in parentheses are \tempo\ reported $1\sigma$ uncertainties.
\end{deluxetable}

%\begin{figure}
%\plotone{lcfluxes1822}
%\caption{Crust cooling curves compared to the observed luminosity. The two models shown have similar energies deposited in the outer crust, but over different portions of the surface, either the whole surface (dashed curve) or 15\% of the area near the magnetic pole (solid curve).}
%\label{fig:cooling}
%\end{figure}

\end{document}